\documentclass[aps, prc, twocolumn,showkeys,showpacs,amsmath,amssymb,superscriptaddress]{revtex4-1}

\usepackage{lineno,hyperref}
\modulolinenumbers[5] 

\usepackage{todonotes}
\usepackage{makeidx}
\usepackage{graphics} 
\usepackage{color}
\usepackage{listings}
\usepackage{bm}
\usepackage{url}
\usepackage{booktabs}
\usepackage{amsmath}
\usepackage{paralist}
\usepackage{amssymb}
\usepackage{amsthm}
\usepackage{subcaption}
\usepackage{algorithm}
\usepackage{algorithmic}
\usepackage{lipsum}

\usepackage[percent]{overpic}

\graphicspath{ {plot/} }

\newcommand*\diff{\mathop{}\!\mathrm{d}}

\begin{document}

%===================================================================================================

\title{ Dissipative hydrodynamics of relativistic shock waves in a Quark Gluon Plasma: 
        comparing and benchmarking alternate numerical methods
      } 
\author{A. Gabbana}
\affiliation{Universit\`a di Ferrara and INFN-Ferrara, I-44122 Ferrara,~Italy}
\affiliation{Bergische Universit\"at Wuppertal, D-42119 Wuppertal,~Germany}
\author{S. Plumari}
\affiliation{Department of Physics and Astronomy, University of Catania, 1-95125 Catania,~Italy}
\affiliation{Laboratori Nazionali del Sud, INFN-LNS, I-95123 Catania,~Italy}
\author{ {G. Galesi} }
\affiliation{Department of Physics and Astronomy, University of Catania, 1-95125 Catania,~Italy}
\affiliation{Laboratori Nazionali del Sud, INFN-LNS, I-95123 Catania,~Italy}
\author{V. Greco}
\affiliation{Department of Physics and Astronomy, University of Catania, 1-95125 Catania,~Italy}
\affiliation{Laboratori Nazionali del Sud, INFN-LNS, I-95123 Catania,~Italy}
\author{D. Simeoni}
\affiliation{Universit\`a di Ferrara and INFN-Ferrara, I-44122 Ferrara,~Italy}
\affiliation{Bergische Universit\"at Wuppertal, D-42119 Wuppertal,~Germany}
\affiliation{University of Cyprus, CY-1678 Nicosia,~Cyprus}
\author{S. Succi}
\affiliation{Center for Life Nano Science @ La Sapienza, Italian Institute of Technology, Viale Regina Elena 295, I-00161 Roma,~Italy}
\affiliation{Istituto Applicazioni del Calcolo, National Research Council of Italy, Via dei Taurini 19, I-00185 Roma,~Italy}
\author{R. Tripiccione}
\affiliation{Universit\`a di Ferrara and INFN-Ferrara, I-44122 Ferrara,~Italy}

\begin{abstract}

  This paper presents numerical cross-comparisons and benchmark results for two different 
  kinetic numerical methods, capable of describing relativistic dissipative fluid dynamics in a wide range of 
  kinematic regimes, typical of relevant physics applications, such as transport phenomena in quark-gluon plasmas. 
  We refer to relativistic lattice Boltzmann versus Montecarlo Test-Particle methods. 
  Lacking any realistic option for accurate validation vis-a-vis experimental data, we check the 
  consistency of our results against established simulation packages available in the literature.
  We successfully cross-compare the results of the two aforementioned numerical approaches for momentum
  integrated quantities like the hydrostatic and dynamical pressure profiles, the collective flow and the heat flux. 
  These results corroborate the confidence on the robustness and correctness of these computational methods 
  and on the accurate calibration of their numerical parameters with respect to the physical transport coefficients. 
 Our numerical results are made available as supplemental material, with the aim of establishing 
  a reference benchmark for other numerical approaches.
\end{abstract}

\maketitle

%===================================================================================================
\section{Introduction}\label{sec:intro}
%===================================================================================================

In its broadest and most general sense, hydrodynamics is the science of collective
behaviour, i.e. the description of the dynamics emerging on top of the rules which
govern the microscopic world.
As a result, hydrodynamic behavior is expected to settle in the presence of a
clearcut separation between the collective degrees of freedom and the
microscopic ones from which they emerge. A safe margin in this respect is about
two orders of magnitude, but the specific threshold at which such emergence
takes place may vary from system to system and in some cases  much less
stringent thresholds may be present. 
Although usually associated with macroscopic motion, distinct hydrodynamic
signatures can be found way deep into the microscopic world, down to the
nanometric scale. In the last two decades relativistic hydrodynamics has
also captured major interest from  the apparently widely separate discipline of
high-energy physics, the main driver being provided by the famous AdS/CFT
theory, which sets an equivalence between d-dimensional field theory and (d+1)
dimensional gravity \cite{maldacena-ijtp-1999}. This fascinating connection has 
opened an active field of modern research on so-called holographic fluids, i.e. strongly
interacting quantum-relativistic fluids supporting the AdS/CFT duality. Among
others, spectacular realizations of hydrodynamic holography have been reported
for electron flows in graphene and quark-gluon plasmas \cite{florkowski-rpp-2018, lucas-jop-2018, romatschke-book-2019}. 
Perhaps, most spectacular of all, recent experiments of the PHENIX collaboration \cite{phenix-natphys-2019}, 
have reported evidence of the ''smallest droplet ever'', namely a droplet of
quark-gluon plasma of the size of a few femtometers!

These manifestations of hydrodynamic behaviour at truly short scales
have spurred a major activity on  the experimental, theoretical and, to a
lesser extent, computational side.
With specific reference to quark-gluon droplets, a major question pertains to
the departures from strict hydrodynamic regimes which take place in small systems
due to partial lack of ''thermalization'' of the initial strongly non-equilibrium
configuration.

The quantitative description of such departures is beyond analytics, and raises
major computational challenges, since a kinetic-theory treatment is
not only computationally very demanding, but also theoretically questionable since
strong-interactions imply supershort-lived quasi-particles, thus undermining the
very premise of kinetic theory.

On the other hand, a macroscopic description of dissipative hydrodynamics poses
further conceptual problems, since it has long been recognized that a naive relativistic 
extension of the Navier-Stokes equations is inconsistent with relativistic invariance, 
implying superluminal propagation, hence non-causal and unstable behavior. 
This can be corrected by resorting to fully-hyperbolic formulations of relativistic hydrodynamics. 
However, while various frameworks have been proposed, the definition of second-order relativistic viscous 
hydrodynamic equations is still debated, with a lot of ongoing research \cite{muronga-prc-2007,muronga-prc-2007b,
betz-epj-2009,el-prc-2010, denicol-prl-2010,betz-epj-2011,denicol-prd-2012,
jaiswal-prc-2013,jaiswal-prc-2013b,jaiswal-prc-2013c,bhalerao-prc-2013,bhalerao-prc-2014,chattopadhyay-prc-2015}.

In this context, approaches based on a mesoscopic description help overcoming some of these problems. 
This approach is useful as, eventually, one may want to study ''beyond-hydrodynamics'' phenomenology;
on the other hand, a consistent description of the hydrodynamic regime requires establishing a link
between the meso-scale and the macroscopic parameters \cite{denicol-prl-2010, 
denicol-prd-2012, molnar-prd-2014, jaiswal-prc-2013b, tsumura-epj-2012, Plumari:2012ep, mendoza-josm-2013, 
florkowski-prc-2015, tsumura-prd-2015, kikuchi-prc-2015, kikuchi-pla-2016, gabbana-pre-2017b,
perciante-josp-2017, ambrus-prc-2018b, gabbana-pre-2019, kurkela-prl-2019}. 
More in general, there is a strong need for reliable numerical tools, as well as numerical benchmarks 
to compare accuracy, stability and performance of these solvers. 

Typical benchmarks used in the validation of relativistic hydrodynamics codes
are the Bjorken flow \cite{bjorken-prd-1983}, Gubser flow \cite{gubser-prd-2010}, 
and the Riemann problem \cite{thompson-jofm-1986,marti-jofm-1994}. The latter is
particularly useful to investigate the robustness of a numerical method due to
the presence of strong discontinuities that give origin to the formation of
shock waves. While the benchmark has an analytic solution only for two limiting
cases, namely for the inviscid and the ballistic regimes,  in the recent past
several studies have analyzed the formation of relativistic shock waves in
viscous QGP matter \cite{bouras-prl-2009,bouras-npa-2009,bouras-prc-2010,bouras-jop-2010}.

To the best of our knowledge, all previous works have been restricted solely to
the study  of the ultra-relativistic regime, for which the appropriate equation
of state (EOS) writes as $\epsilon = 3 P$.
In this work we consider instead a more general EOS, accounting for massive particles:
\begin{equation}\label{eq:relativistic-eos-3D}
\begin{array}{rll}
  \epsilon &=& P \left( 3 + \zeta \frac{K_{1}(\zeta)}{K_{2}(\zeta)} \right) \quad , \\
         P &=& n k_{\rm B} T \quad ,
\end{array}
\end{equation}
where the relativistic parameter $\zeta = m c^2 / k_B T$, defined as the ratio 
between the rest mass energy $mc^2$ and the thermal energy $k_B T$, physically characterizes the 
kinematic regime of the macroscopic fluid, with $\zeta \rightarrow 0$ 
in the ultra-relativistic regime and $\zeta \rightarrow \infty$ in the non-relativistic one.
This approach allows to test EOS quite similar to the one coming from lattice QCD calculations,
even if a very precise description of the most recent EoS requires to consider a temperature dependent 
mass as shown in Ref. \cite{plumari-prd-2011}. Such an approach has been already implemented within the relativistic Boltzmann 
approach in \cite{plumari-plb-2010} and discussed also to derive relativistic viscous hydrodynamics \cite{tinti-prd-2017}.

We apply this benchmark with constant mass and compare two different computational methods, and present numerical results exploring a 
wide range of parameters. We start by replicating results available in the literature in
the ultra-relativistic limit and then move on to the study of fluids of massive particles,
relevant to the QGP framework. 
The methods we consider both share a kinetic approach at the mesoscale level, but differ significantly in their numerical formulation, 
namely i) a Relativistic Lattice Boltzmann (RLBM) approach, based on the relaxation time approximation, 
and ii) a Monte Carlo-enabled solution of the full kernel of the Boltzmann equation based on the Test-Particle-Method (RBM-TP).
They are cross-validated for small values of the ratio between shear viscosity 
and the entropy density ($\eta / s < 0.2$ ), corresponding to a hydrodynamic regime
where the Knudsen number is $\rm{Kn} \ll 1$. We show that the two solvers provide results in very good agreement,
by analysing the profile of momentum integrated macroscopic quantities, as well as the non-equilibrium contributions
to the moments of the particle distribution function 
\cite{gan-sm-2015,montessori-pre-2015}.

We also show that, as expected, the RLBM approach fails when $\eta/s$ is significantly large (ballistic limit)  
while successfully captures the physics features of flows at very low shear viscosity. The Monte Carlo approach, RBM-TP,  
is able to get solutions in agreement with RLBM even at very low viscosity, $\eta/s \simeq 0.05$, 
being able to naturally describe the evolution also for systems
at large viscosity up to the ballistic limit $\eta/s \rightarrow \infty$.

The numerical results presented in this work are made available as supplementary material with the aim of promoting further future comparisons \cite{SOM}. 

%===================================================================================================
%===================================================================================================
\begin{figure*}[htb]
  \centering
  \includegraphics[width=0.95\textwidth]{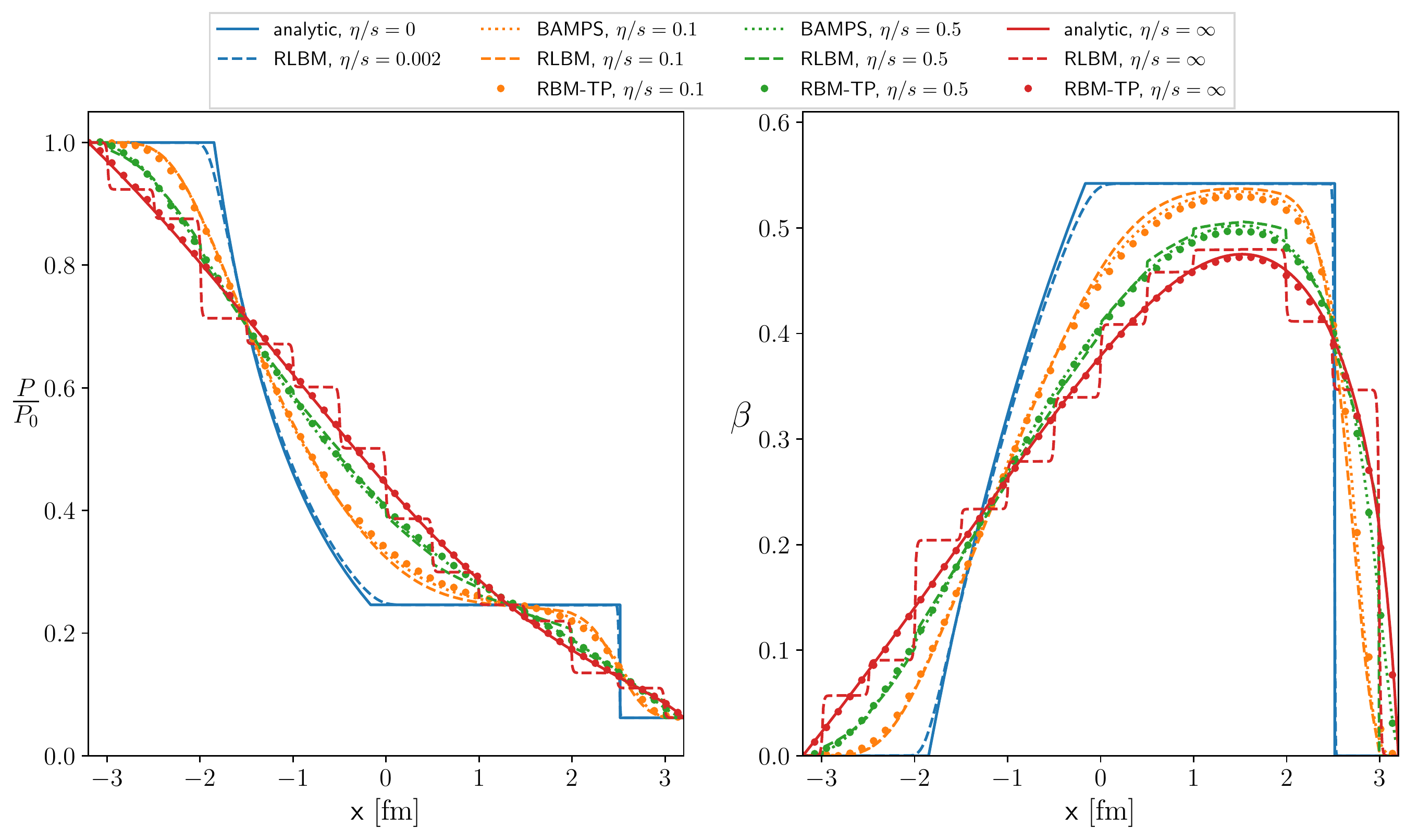}
  \caption{ Riemann problem for a gas of massless particles for various viscous regimes. 
            The left panel shows the pressure profile and the right one shows $\beta = |U^i|/U^0$, at $t = 3.2~\rm{fm/c}$. 
            Solid lines are analytic solutions, dotted lines are numerical results 
            obtained by BAMPS, dashed lines refer to RLBM, and dots to RBM-TP. 
            Different colors (see online version) represent different viscous regimes, 
            with blue for the inviscid limit, orange for $\eta/s = 0.1$, green for $\eta/s = 0.5$, 
            and red for free-streaming.
          }\label{fig:cmp-bamps}
\end{figure*}
%===================================================================================================
%===================================================================================================

In what follows we adopt natural units, for which $\hbar = c = k_{\rm B} = 1$, and 
a flat space-time described by the metric tensor $\eta = (1, -1, -1, -1)$.

%===================================================================================================
\section{Model equations}\label{sec:hydro}
%===================================================================================================

The kinetic description of a relativistic gas is based on the particle distribution function
$f( (x^{\alpha}), (p^{\alpha}) )$, depending on space-time coordinates $\left( x^{\alpha} \right) = \left( t, \bm{x} \right)$ 
and momenta $\left( p^{\alpha} \right) = \left( p^0, \bm{p} \right) = \left( \sqrt{\bm{p}^2 + m^2}, \bm{p} \right)$, 
with $\alpha = 0, 1, 2, 3$. The space-time evolution of $f( (x^{\alpha}), (p^{\alpha}) )$ is
governed by the relativistic Boltzmann equation which, in the absence of external forces, writes as
\begin{equation}\label{eq:relativistic-boltzmann}
  p^{\alpha} \frac{\partial f}{\partial x^{\alpha}}
  =  
  C[f] \quad ,
\end{equation}
$C[f]$ being the collisional operator.

Macroscopic quantities are defined from the moments of the distribution functions.
The first moment is the particle four-current:
\begin{equation}\label{eq:particle-flow}
  N^{\alpha} = \int f p^{\alpha} \frac{\diff^3 p}{p_0} \quad ,
\end{equation}
while the second moment defines the energy-momentum tensor:
\begin{equation}\label{eq:energy-momentum-tensor}
  T^{\alpha \beta} = \int f p^{\alpha} p^{\beta} \frac{\diff^3 p}{p_0} \quad .
\end{equation}
The balance equations of the particle four-current and of the energy-momentum 
tensor deliver the following conservation equations:
\begin{equation}\label{eq:conservation-eqs}
  \begin{aligned}  
    \partial_{\alpha} N^{\alpha}       &= 0 \quad , \\
    \partial_{\alpha} T^{\alpha \beta} &= 0 \quad .
  \end{aligned}
\end{equation}

These equations are purely formal until a specific form for $N^{\alpha}$ and $T^{\alpha \beta}$ is specified.
Following Landau and Lifshitz \cite{landau-book-1987} one has
\begin{align}
  N^{\alpha}       &= n U^{\alpha}  - \frac{n}{P + \epsilon} q^{\alpha} \quad                                                  , \label{eq:ll-decomposition-o1} \\
  T^{\alpha \beta} &= \epsilon U^{\alpha} U^{\beta} - \left( P + \varpi \right) \Delta^{\alpha \beta} + \pi^{< \alpha \beta >} , \label{eq:ll-decomposition-o2}  %\quad ; 
\end{align}
with $\epsilon$ the energy density, $P$ the hydrostatic pressure, $q^{\alpha}$ 
the heat flux, $\pi^{<\alpha \beta>}$ the pressure deviator, $\varpi$ the dynamic pressure,
and $\Delta^{\alpha \beta} = \eta^{\alpha \beta} - U^{\alpha} U^{\beta}$ the (Minkowski-)orthogonal projector  
to the fluid velocity $U^{\alpha}$; the latter, in the Landau frame, is defined as $T^{\alpha \beta} U_{\beta} = \epsilon U^{\alpha}$.

%===================================================================================================
\section{Numerical Methods}\label{sec:methods}
%===================================================================================================
In this section, we provide a brief account of the two numerical methods used in this work, 
the Relativistic Lattice Boltzmann Method and the Montecarlo Test-Particle Method.

%===================================================================================================
\subsection{Relativistic Lattice Boltzmann Method}\label{sec:rlbm}
%===================================================================================================
The first computational method used in this comparison is a relativistic lattice Boltzmann method.
This method \cite{mendoza-prl-2010,romatschke-prc-2011,gabbana-pre-2017} 
is a computationally efficient approach to dissipative relativistic hydrodynamics. 
One key advantage over other relativistic hydrodynamic solvers is that, being based on a mesoscale approach, 
the emergence of viscous effects does not break relativistic invariance and causality, 
since space and time are treated on the same footing, i.e. both via first order derivatives (hyperbolic formulation).

This numerical method solves a minimal version of Eq.~\ref{eq:relativistic-boltzmann}, where the 
microscopic momentum vector is discretized on a Cartesian grid, and with the collisional 
operator replaced by the Anderson-Witting single relaxation time approximation \cite{anderson-witting-ph-1974a,anderson-witting-ph-1974b}:
\begin{equation}\label{eq:anderson-witting}
  C[f] = - \frac{U^{\alpha} p_{\alpha}}{\tau} \left( f - f^{\rm eq} \right) \quad .
\end{equation}
In the above, $\tau$ is the relaxation proper-time and $f^{\rm eq}$ is the equilibrium 
distribution for which we consider the Maxwell-J\"uttner statistics:
\begin{equation}\label{eq:MJ-edf}
  f^{\rm eq} = \frac{n}{4 \pi T^3 \zeta^2 K_2(\zeta)} \exp{\left(- \frac{U^{\alpha} p_{\alpha}}{T} \right)} \quad ;
\end{equation}
here and in the following $K_i(\zeta)$ is the modified Bessel function of the second kind of order $i$.
The connection between the microscopic parameter $\tau$ and the macroscopic equations is
given by the transport coefficients, for which we take into account the analytic expressions
resulting from the first order Chapman-Enskog expansion \cite{cercignani-book-2002}. 
Relevant for the present study is the shear viscosity $\eta$:
\begin{equation}\label{eq:eta-ce}
  \eta = P \tau \frac{\zeta}{15} \left(3 \frac{K_3(\zeta)}{K_2(\zeta)}  - \zeta + \zeta^2 \frac{K_1(\zeta)}{K_2(\zeta)} - \zeta^2 \frac{Ki_{1}}{K_2}\right) \quad ,
\end{equation}
with $Ki_{1}$ the Bickley-Naylor function
\begin{equation*}
  Ki_{\alpha} =  \int_0^{\infty} e^{-\zeta \cosh (t)} \left( \cosh (t) \right)^{- \alpha} \diff t \quad .
\end{equation*}

%===================================================================================================
\subsection{The Test Particle Method}\label{sec:catania}
%===================================================================================================

The second computational approach considered in this paper 
also belongs to the realm of kinetic transport theory.
We use a relativistic transport code developed to perform studies of the dynamics 
of heavy-ion collisions at both RHIC and LHC energies \cite{Plumari:2012ep,Ruggieri:2013bda,Scardina:2013nua,Ruggieri:2013ova,Puglisi:2014sha,Plumari:2015sia,Plumari:2015cfa,Scardina:2017ipo,Plumari:2019gwq}.

Recently, the code has been extended to the solution of equation Eq.~\ref{eq:relativistic-boltzmann} for massive particles, which allows 
to simulate a fluid with an EOS that can be close to the recent lQCD calculations \cite{Plumari:2019gwq}. 
In this work, Eq.~\ref{eq:relativistic-eos-3D} has been used.

In this work we consider only $2 \leftrightarrow 2$ collision processes, 
which give rise to a collisional operator $C[f]$ of the form:
\begin{equation*}
  C[f] = \int \frac{d^3{p}_{2}}{E_{2}} d\Omega  \frac{1}{2}\sqrt{s(s-4m^2)} \,\, \frac{d\sigma(s,\Theta)}{d\Omega} (f_1^{\prime}f_2^{\prime} - f f_2)  \quad ;
\end{equation*}
$f_i=f(p_i)$ ($f_i'=f(p_i')$) are the distribution functions of the outgoing (ingoing) particles, 
$s = (p_1+p_2)^2$ and $\sigma(s,\Theta)$ is the differential cross section which 
is related to the total cross section by $\sigma_{\rm tot}=\int d\Omega\, \frac{d\sigma(s,\Theta)}{d\Omega}$. 
The numerical solution of the transport equation is obtained by using the test particle method, a popular option
in many transport calculations \cite{Bertsch:1988ik,Cassing:1990dr,Buss:2011mx}.
In this method, the phase-space distribution function is sampled by mean 
of a large number of so-called test particles.
In fact, it can be shown, that the phase space distribution given by a collection of 
point-like test particles is a solution of the Boltzmann equation, provided the positions and momenta of the test particles 
obey the relativistic Hamilton equations \cite{Wong:1982zzb, Ko:1987gp}.

The numerical implementation of the collision integral is based on the so-called stochastic method \cite{Xu:2004mz}
that has proven capable of describing efficiently also the ultra-relativistic limit, avoiding the
issue of causality induced by a geometrical interpretation of the collision integral.

%===================================================================================================
%===================================================================================================
\begin{figure}[htb]
  \centering
  \includegraphics[width=0.99\columnwidth]{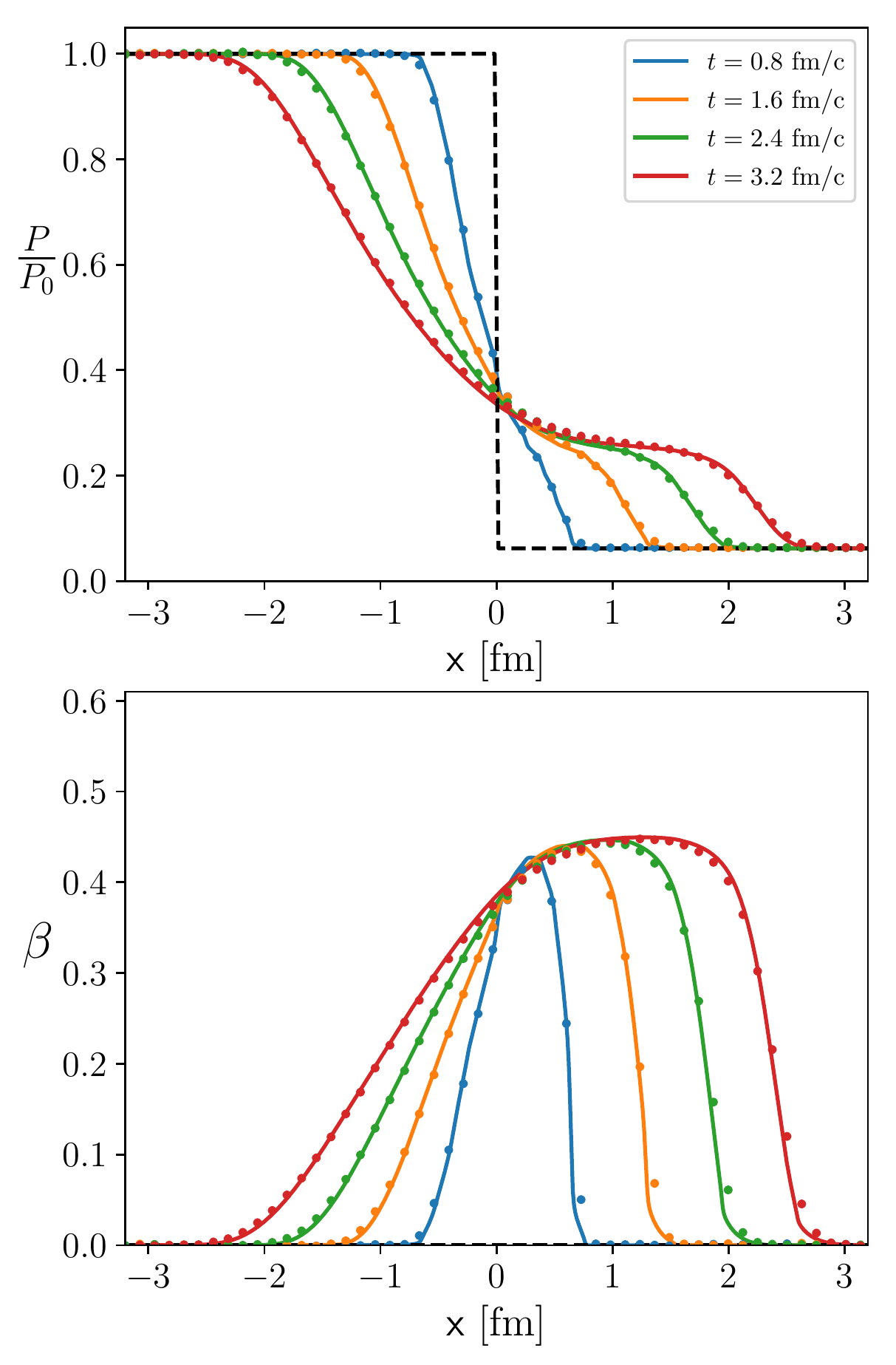}
  \caption{ Riemann problem for a gas of particles with rest mass $m=0.8~\rm{GeV}$, at $\eta /s = 0.1$.
            The top panel shows the pressure profile and the bottom one shows $\beta = |U^i|/U^0$, 
            at $t = 0.8, 1.6, 2.4$ and $3.2~\rm{fm/c}$. The initial conditions are represented by
            black dotted lines. Continuous lines are results obtained by RLBM, 
            while dots are results by RBM-TP.
          }\label{fig:m08-etas01}
\end{figure}
%===================================================================================================
%===================================================================================================

The transport code permits to study the dependence of physical observables on microscopical processes fixed by 
matrix elements or cross section. 
A novel approach is based on the idea of gauging the collision kernel $C[f]$ to a desired $\eta/s$ ratio by an 
effective cross section $\sigma_{\rm tot}$ \cite{Plumari:2012ep,Ferini:2008he,Ruggieri:2013ova} .
Such an  approach was inspired by the success of the hydrodynamical approach to describe experimental 
data \cite{Gardim:2012yp,Schenke:2011bn,Gale:2012rq} and permits to employ a Boltzmann transport equation
in regimes where $\eta/s$ (or equivalently the scattering relaxation time $\tau \sim 1/\sigma\rho$) is 
very low  \cite{Huovinen:2008te,Molnar:2008xj,Plumari:2015sia}. 

The expression given by the first order Chapman-Enskog expansion (Eq.~\ref{eq:eta-ce}), is used for the shear viscosity,
with the relaxation time defined as $\tau = R^{-1}=\frac{1}{n \langle \sigma_{\rm tr} v_{\rm rel} \rangle}$, 
with $\sigma_{tr}$ the transport cross section, $v_{\rm rel}$ is the relative velocity of the two incoming particles. 
In Ref. \cite{Plumari:2012ep} it has been checked through the Green-Kubo correlator that employing
Eq.~\ref{eq:eta-ce} generates a fluid with the desired value of $\eta/s$.

For the massive case $R$ is given by the following expression:
\begin{equation}\label{eq:rate}
  R = n\langle \sigma_{\rm tr} v_{\rm rel} \rangle 
    = n \frac{\beta}{4} \frac{\int_{\sqrt{s_0}}^{\infty} \, \, d\sqrt{s} \,\, \lambda(s) \, \sigma_{\rm tr}(s) \, K_1(\beta \sqrt{s})}{[m^2 K_2(\beta m)]^2} \quad ,
\end{equation}
where $\sqrt{s_0}=2 m$ and $\lambda(s)= s [ s- (2 m)^2 ]$.
Note that for massless particles and constant isotropic cross-section the above formula 
reduces to $R \to n \frac{2}{3}\sigma_{\rm tot}$. Finally, Eq.~\ref{eq:eta-ce} and Eq.~\ref{eq:rate} provide the formula 
for the normalization of the cross section in each cell in order to keep fixed $\eta/s$.

%===================================================================================================
\section{Numerical Results}\label{sec:numerics}
%===================================================================================================

%===================================================================================================
%===================================================================================================
\begin{figure}[htb]
  \centering
  \includegraphics[width=0.99\columnwidth]{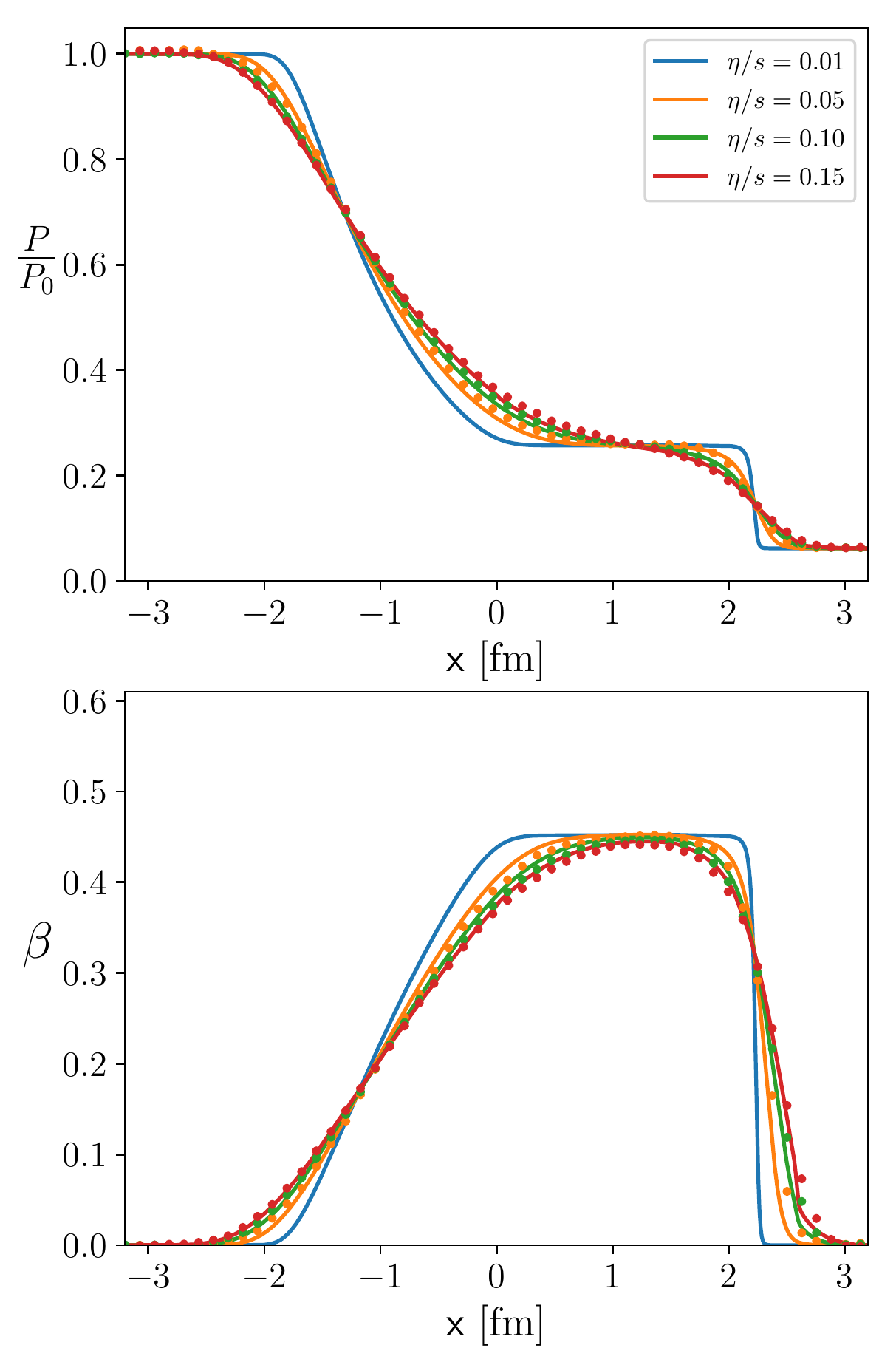}
  \caption{ Riemann problem for a gas of particles with rest mass $m=0.8~\rm{GeV}$, for various viscous regimes.
            The top panel shows the pressure profile and the bottom one shows $\beta = |U^i|/U^0$, at $t = 3.2~\rm{fm/c}$. 
            Lines are results obtained by RLBM, while dots are results provided by the RBM-TP.
          }\label{fig:fixed-mass}
\end{figure}
%===================================================================================================
%===================================================================================================

For our numerical analysis, we begin by considering the relativistic Riemann problem.
This problem describes a tube filled with a gas which initially is 
in two different states on the two sides of a membrane placed at $x = 0$.
As a result, the macroscopic quantities describing the fluid present a discontinuity at the membrane.
Once the membrane is removed, the discontinuities decay producing shock/rarefaction waves,
depending on the initial configuration chosen for the two different chambers.
For simplicity we assume the flow to be homogeneous in the transverse directions.

Analytical solutions for this problem are available only for the massless case, and for extreme values of 
the ratio $\eta / s$, i.e. in the inviscid limit ($\eta / s \rightarrow 0$) \cite{thompson-jofm-1986},
and in the free-streaming limit ($\eta / s \rightarrow \infty$) \cite{greiner-prc-1996}.

We use the same initial conditions as in \cite{bouras-npa-2009}, with a pressure jump
defined as $P_0 = 5.43~\rm{GeV~fm^{-3}}$ for $x < 0$ and $P_1 = 0.33~\rm{GeV~fm^{-3}}$ for $x > 0$.
Initial values for the temperature are respectively $T_0 = 400~\rm{MeV}$ and $T_1 = 200~\rm{MeV}$.

We perform simulations at constant values of $\eta / s$. 
The calculation of $\eta$ follows from the discussion in the previous section, 
while the entropy density is approximated using \cite{cercignani-book-2002}
\begin{equation}\label{eq:entrophy-density}
  s = n \left( \zeta \frac{K_3(\zeta)}{K_2(\zeta)} - \ln{(\frac{n}{n^{\rm eq}})} \right) \quad ;
\end{equation}
we use the following expression for the equilibrium density
\begin{equation}\label{eq:equilibrium-density}
  n^{\rm eq} = d_{\rm G} \frac{T^3}{2 \pi^2} \zeta^2 K_2(\zeta) \quad ,
\end{equation}
with $d_{\rm G} = 16$ the degeneracy factor of gluons. 

Combining Eq.~\ref{eq:eta-ce}, ~\ref{eq:entrophy-density} and ~\ref{eq:equilibrium-density}
it is then possible to define the relaxation time required to keep the ratio $\eta / s$ 
constant to a desired value $k$. As an example, the expression in the ultra-relativistic limit reads as:
\begin{equation}
  \tau = k \frac{5}{4 T} \left[ 4 - \log{\left(\frac{\pi^2 n}{d_{\rm G} T^3} \right)} \right] \quad .
\end{equation}
In RLBM simulations, the above equation is used to locally adjust the relaxation time,
while in RBM-TP simulations it is used in combination with Eq.~\ref{eq:rate} to calculate 
the corresponding value of the transport cross-section.
Table~\ref{tab:tau-values} lists the values of the relaxation time at the initial step of the simulation corresponding to $\eta / s = 0.1$, for all 
values of the particle rest-mass $m$ considered in this work.

\begin{table}[htb]
\centering
\resizebox{0.6\columnwidth}{!}{
\begin{tabular}{|c|c|c|}
  \hline 
                   & \multicolumn{2}{c|}{$\eta/s = 0.1$} \\
  \hline 
    $m~[\rm{GeV}]$ & $\tau_{\rm 0}~[\rm{fm/c}]$ & $\tau_{\rm 1}~[\rm{fm/c}]$ \\
  \hline                                    
    $0  $ & 0.246 & 0.496 \\
    $0.8$ & 0.258 & 0.551 \\
    $2  $ & 0.281 & 0.621 \\
    $4  $ & 0.309 & 0.692 \\
  \hline 
\end{tabular}
}
\caption{
	Values of the relaxation times corresponding to $\eta/s = 0.1$. For all values of the particle rest-mass $m$,
	we list the relaxation times corresponding to
	the initial conditions on the left ($\tau_0$) and on the
	right hand side ($\tau_1$) of the discontinuity.
        }\label{tab:tau-values}
\end{table}

As a warm-up exercise, we start by reproducing the results of previous studies in the ultra-relativistic regime.
In particular, we compare against the Parton cascade Boltzmann Approach to Multi-Parton Scatterings (BAMPS) \cite{xu-prc-2007}, 
which numerically solves the Boltzmann equation using a Monte-Carlo approach.
In Fig.~\ref{fig:cmp-bamps}, we show that both the methods correctly 
reproduce the results provided by BAMPS at $\eta / s = 0.1$. RLBM also gives a good 
approximation to the analytical solution in the inviscid limit; in this case the test-particle 
methods cannot be applied since for $\eta / s \rightarrow 0$ the cross section becomes 
unphysically large leading to numerical instabilities. On the other hand for large values of $\eta / s$,
the hydrodynamic approach becomes questionable, as we transit towards a ballistic regime.
Although at $\eta / s = 0.5$ RLBM still manages to capture the qualitative behaviour of the flow, 
in this regime the results provided by the RBM-TP are more reliable and in better
agreement with BAMPS. 
It is however important to note that even at the very low $\eta/s=0.05$, the RBM-TP 
is  able to describe  the dynamical evolution in excellent agreement with RLBM.
Finally, in the free-streaming limit, the test-particle method reproduces correctly
the analytic solution, while an unphysical ``staircase'' effect is observed in the profiles produced by RLBM;
it has been shown that higher order schemes can cure this issue \cite{ambrus-prc-2018}.
We remark that the RLBM simulations carry a systematic error due to the truncation of the
higher order moments of the particle distribution (see \cite{gabbana-pre-2017} for details),
whereas a statistical error is  inherently associated with RBM-TP. 
However in this and in all other figures in this paper, error bars are not shown, since we average over a sufficiently 
large number of events in order to keep statistical errors well below $1 \%$ for the macroscopic observables.

In the following, we take into consideration fluids consisting of massive particles 
and restrict our analysis to values of $\eta /s < 0.2$, 
corresponding to a hydrodynamic regime for which we have observed a 
good agreement between the two methods. We recall that such a regime is also the one of interest
for the study of the quark-gluon plasma (QGP) in ultra-relativistic collisions \cite{Heinz:2013th,Gale:2013da}.

%===================================================================================================
%===================================================================================================
\begin{figure}[htb]
  \centering
  \includegraphics[width=0.99\columnwidth]{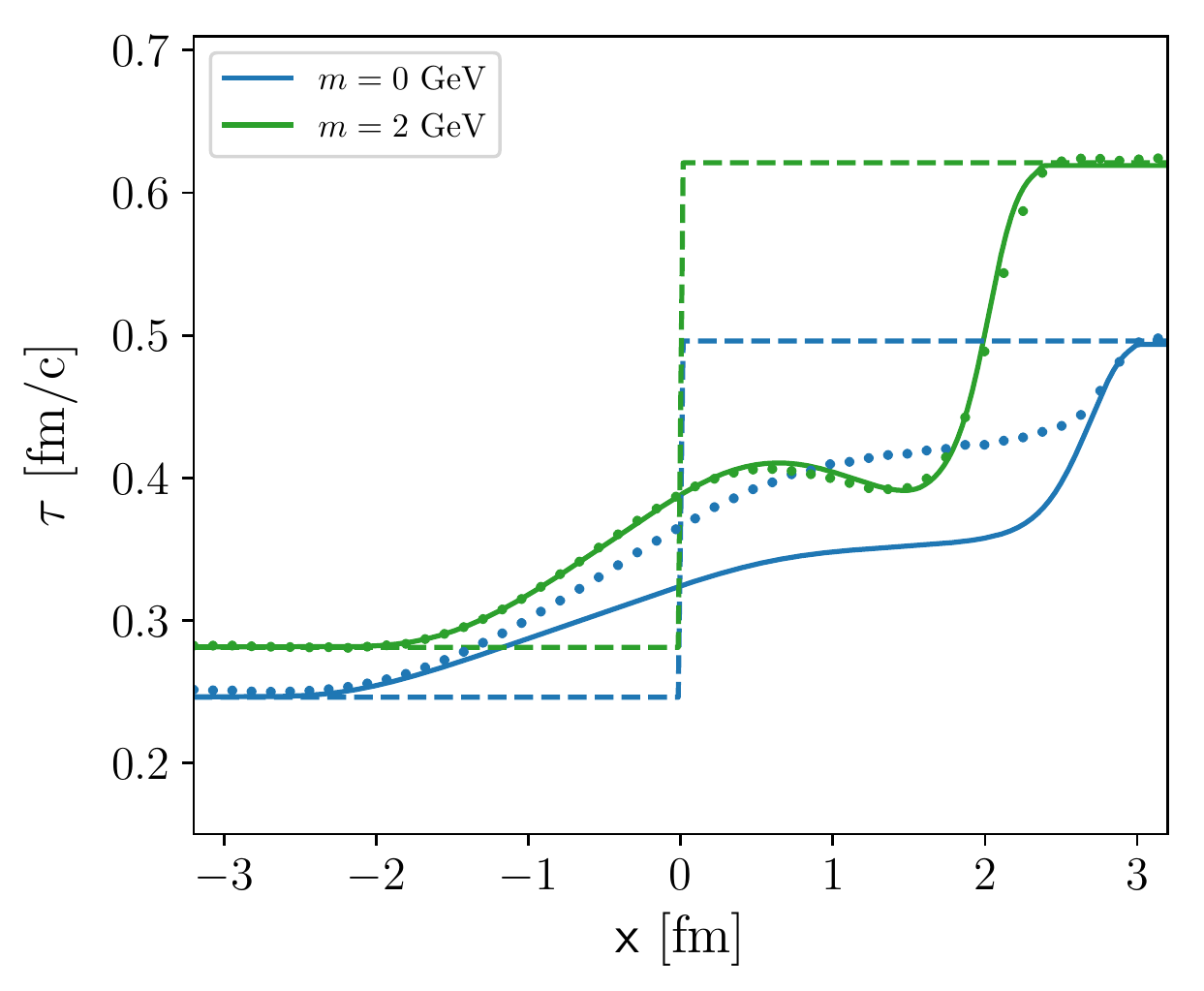}
  \caption{Local viscous relaxation time $\tau$ for the case of $\eta/s=0.1$ 
  at time t = 0 (dashed lines) and t = 3.2  fm/c for m=0 and m=2 GeV. 
  The solid lines are the values for the RLBM and the dots are the values of the RBM-TP.
  }\label{fig:tau}
\end{figure}
%===================================================================================================
%===================================================================================================

In Fig.~\ref{fig:m08-etas01}, we show the time evolution of the pressure
and velocity profiles for a specific case with $m = 0.8~\rm{GeV}$
and $\eta/s = 0.1$. We observe an excellent agreement in the dynamics 
predicted by the two solvers; in particular, we remark that the two
cases compare well not only at times $t$ much longer than the
relaxation time $\tau$, but also in the short-term regime, $t \sim \tau$.

In Fig. \ref{fig:tau} we also show the space distribution of the relaxation time $\tau(x)$ in the
two approaches for the case of $\eta/s=0.1$. 
It is seen that, even if the  initial values, dashed lines, are the same according to
 Eq. (\ref{eq:eta-ce}), the local evolution in the region of the shock wave
 exhibits different values for RLBM (solid lines) and RBM-TP (dots) 
for the massless case, a discrepancy that nearly disappears for the massive case m=2 GeV.
This can be attributed to the different collision kernels: the Anderson-Witting in RLBM and the full 
Boltzmann in RBM-TP; however, as shown in all other results, such difference does not
lead to any appreciable difference of macroscopic quantities, like the hydrostatic pressure 
or the collective flow. 
Some difference is observed instead in Fig. \ref{fig:tensors},  which reports non-equilibrium 
quantities, such as the dynamical pressure or heat flux. Such differences
also tend to vanish for the very massive case, as we are going to discuss
at the end of this section.

%===================================================================================================
%===================================================================================================
\begin{figure}[htb]
  \centering
  \includegraphics[width=0.99\columnwidth]{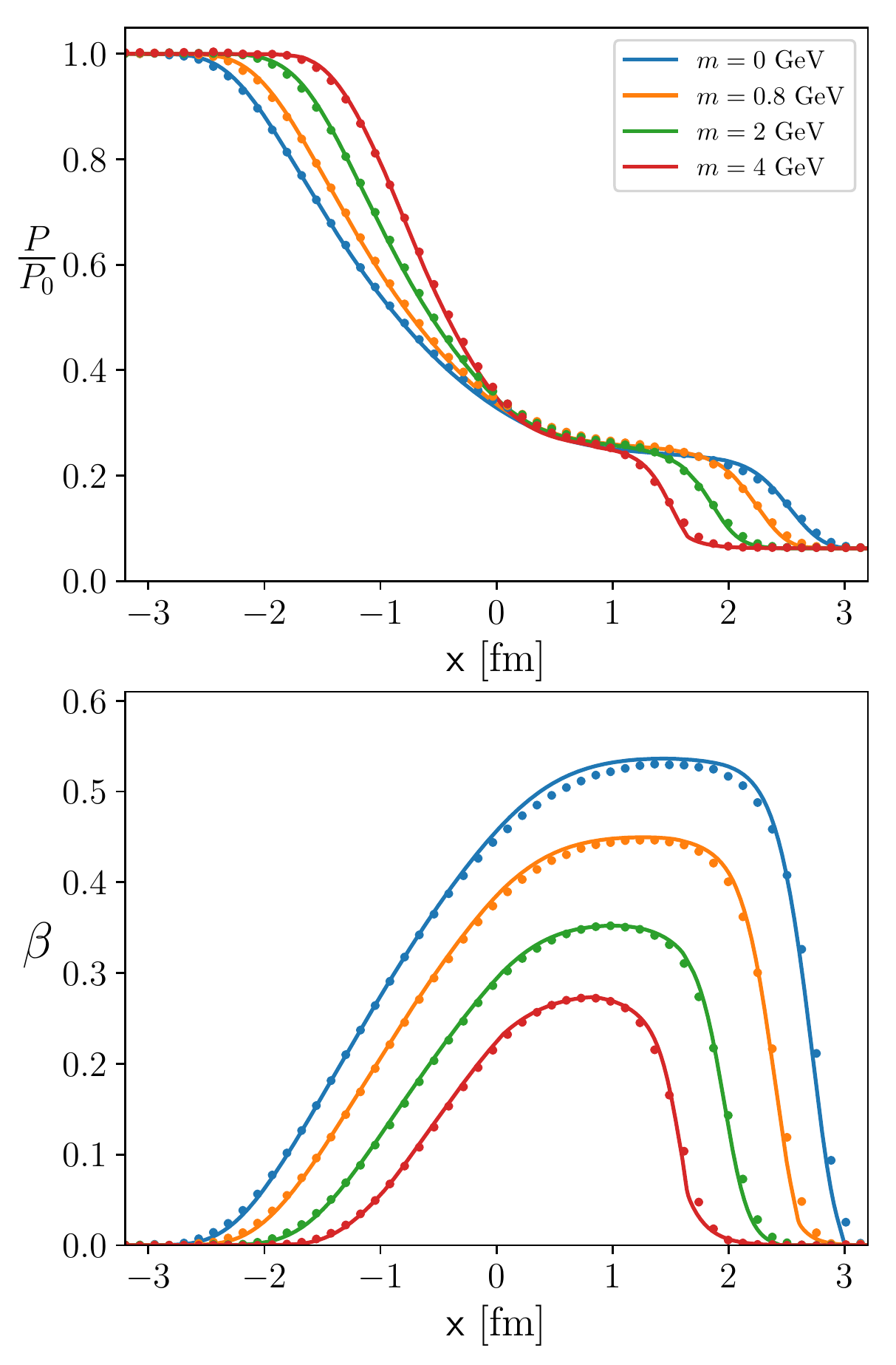}
  \caption{ Riemann problem for a gas of particles with rest mass respectively $0, 0.8, 2$ and $4~\rm{GeV}$.
            All simulations are at $\eta / s = 0.1$.
            The top panel shows the pressure profile and the bottom one shows $\beta = |U^i|/U^0$,
            at $t = 3.2~\rm{fm/c}$. Lines are results obtained by RLBM, dots are results provided by RBM-TP.
          }\label{fig:fixed-etas}
\end{figure} 
\begin{figure*}[tbh]
  \centering
  \includegraphics[width=0.99\textwidth]{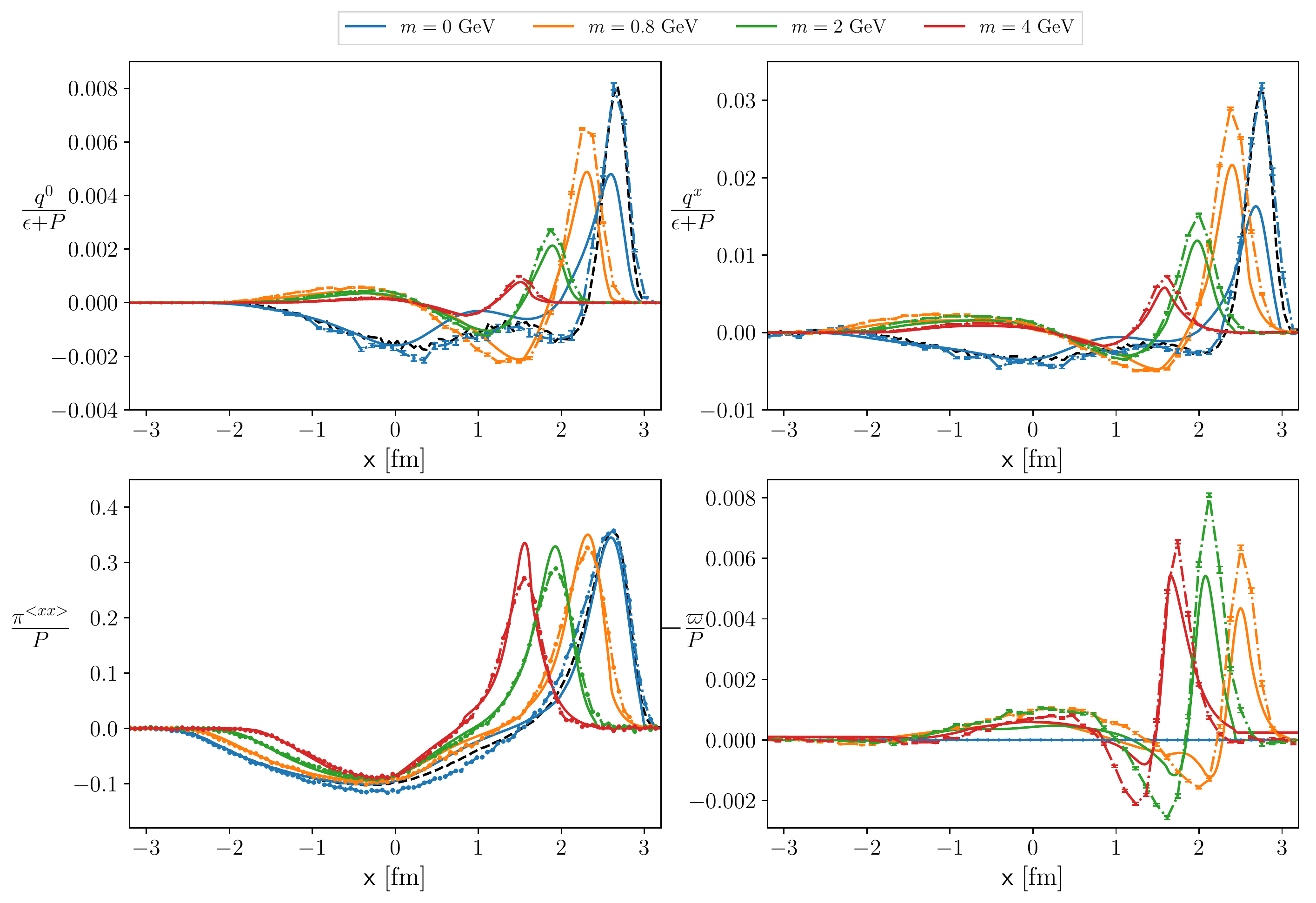}
  \caption{ Analysis of quantities related to the four-flow and the energy-momentum tensors for the same simulation setup of Fig.~\ref{fig:fixed-etas}.
            Top    left: time component of the heat-flux.
            Top    right: spatial component of the heat-flux.
            Bottom left: spatial component of the pressure deviator.
            Bottom right: dynamic pressure.
            Lines are results obtained by RLBM, while dashed points are the results provided by RBM-TP. Black dashed lines are 
            results provided by BAMPS in the ultra-relativistic limit.
          }\label{fig:tensors}
\end{figure*}
%===================================================================================================
%===================================================================================================

In Fig.~\ref{fig:fixed-mass}, we show the results obtained considering 
again a rest mass $m = 0.8~\rm{GeV}$ but for different viscous regimes.
The two methods are in good agreement, with only slight differences
observed in the proximity of the shock-wave front. The curves for RBM-TP at $\eta/s = 0.01$
are not shown since, at such low viscosity, the methods becomes numerically unstable
as the cross section and hence the computational time diverges as $\eta/s \rightarrow 0$.
However it has been shown in Ref. \citep{Plumari:2015sia} that a linear extrapolation
in $1/\sigma$ for $\sigma \rightarrow \infty $ provides the correct pattern even for ideal hydrodynamics.

In Fig.~\ref{fig:fixed-etas} we fix $\eta /s = 0.1$ and compare the results
obtained for fluids of particles with rest mass of $0, 0.8, 2$ and $4~\rm{GeV}$. 
Once again, both computational methods yield the same numerical results, 
which is remarkable given that no parameter fitting is performed in this analysis 
(apart from requirement to keep $\eta/s$ constant).

Furthermore, it is interesting to analyse the non-equilibrium contributions to the 
four-flow tensor $N^{\alpha}$ and to the energy-momentum tensor $T^{\alpha \beta}$.

In the top panel in Fig.~\ref{fig:tensors}, we report the time and the spatial component
of the heat flux, calculated via
\begin{equation}
  q^{\alpha} = \frac{P + \epsilon}{n} \left(n U^{\alpha} - N^{\alpha} \right) \quad .
\end{equation}
The time component of $q^{\alpha}$ is expected to vanish for non-relativistic flows.
We can observe that as the rest mass is increased, the fluid moves at slower speeds
and the correspondent peak in $q^{0}$ reduces accordingly.
We observe significant discrepancies at the peak values of both the time and spatial components
of the heat flux in the mass-less case. RBM-TP closely follows the results of BAMPS,
while RLBM seems to be in better agreement with other hydrodynamic codes, like vSHASTA \cite{bouras-prc-2010}
or other lattice Boltzmann approaches (see e.g. Fig.~6 in \cite{ambrus-prc-2018}).
On the other hand, when considering massive particles the results seem to be in good agreement.
We remark that although calibrated to reproduce the same first order coefficient for the shear viscosity, 
the two methods approximate higher orders in a different way. It has been shown \cite{bouras-prc-2010}
that higher order terms in Knudsen number play a relevant role in the heat flow.

Next, we take into consideration the non-equilibrium part of energy-momentum tensor, in particular
the pressure deviator, defined as
\begin{equation}
  % \pi^{<xx>} = 2\eta \nabla^{<x} U^{x>}
  \pi^{<\alpha \beta>} 
  = 
  \eta \left(                \Delta^{\alpha}_{\gamma} \Delta^{\beta}_{\delta} + 
                             \Delta^{\alpha}_{\delta} \Delta^{\beta}_{\gamma}  
              - \frac{1}{3}  \Delta^{\alpha \beta} \Delta_{\gamma \delta} 
      \right) \nabla^{\gamma} U^{\delta}
\end{equation}
with the shorthand notation
\begin{equation}
\begin{array}{lcl} 
  \nabla^{\alpha}         & = & \Delta^{\alpha \beta} \partial_{\beta}       \quad ,\\
  \Delta^{\alpha}_{\beta} & = & \Delta^{\alpha \gamma} \Delta_{\gamma \beta} \quad .
\end{array} 
\end{equation}

In the bottom left panel of Fig.~\ref{fig:tensors} we show $\pi^{<xx>}$, the spatial component of the pressure deviator.
The two methods provide results in good agreement, both reproducing the results of BAMPS in the ultra-relativistic case,
with slight differences in the massive case at the discontinuity point.

Particularly relevant to the study of QGP is the analysis of effects related to the bulk viscosity, with
potential implications for dark matter \cite{dou-aa-2011,velten-ijgmmp-2014,gagnon-jcap-2011, Ryu:2015vwa}.

The bulk viscosity enters the non-equilibrium contribution to the energy-momentum tensor,
and in particular is proportional to the dynamic pressure:
\begin{equation}
  \varpi = - \mu \nabla_{\alpha} U^{\alpha} \quad .
\end{equation}
We calculate the dynamic pressure from simulations by applying the projector $\Delta_{\alpha \beta}$ 
to Eq.~\ref{eq:energy-momentum-tensor}, which yields:
\begin{equation}
  \varpi = \frac{1}{3} \left( \epsilon - T^{\alpha}_{\alpha} \right) - P .
\end{equation}
Measurements of $\varpi$ are shown in the bottom right panel of Fig.~\ref{fig:tensors}.
This quantity is exactly zero for the massless case, consistent with the fact that a 
ultra-relativistic gas has no bulk viscosity. 
Bulk effects are more prominent for the case $m = 2~\rm{GeV}$, 
which corresponds to $\zeta \approx 5$, consistently with analytical calculations \cite{cercignani-book-2002}.
Differences in the dynamical pressure can have some relevance for the phenomenology of the QGP physics 
because they moderately affect the average momentum of the spectra
and the build-up of azimuthal anisotropic collective flows \cite{Ryu:2015vwa,Ryu:2017qzn}.

In general, we can see that for the heat flux, $q^\alpha$ , and for the dynamical pressure $\varpi$ 
larger differences between the two methods, mainly in the region of the peak.
This is not surprising, because such quantities focus on the details of the non-equilibrium dynamics
and the simplified collisional kernel in RLBM may induce a slightly different non-equilibrium dynamics
with respect to RBM-TP. However, this does not lead to any significant difference in
the dynamic evolution of global quantities,  like the pressure and the collective flow, as shown in all the previous section. 
It is also worth of note that the differences
decrease with increasing mass, corresponding to a less rapid dynamics, and that the differences are larger 
in the same region where the $\tau (x)$, shown in Fig. \ref{fig:tau}, exhibits the largest departure between the two approaches.
In any case, it should be observed that the differences between the two methods in $\varpi$ are anyway
smaller than $0.5 \%$, with respect to the hydrostatic pressure.

We end this section by stressing again that our results, while conceptually based on a Boltzmann mesoscopic description, result 
from two fully different computational approaches; each computational method has its own parameters describing the mesoscale dynamics, but 
in both cases, these values can be derived from transport coefficients at the macroscopic scale. 
The two algorithms provide correct results in different ranges of the transport coefficients: RLBM works correctly in the range $\eta/s \rightarrow 0$  to
$\eta/s \simeq 0.5$, while RBM-TP works correctly for $\eta/s \ge 0.05$
up to the ballistic limit of $\eta/s \rightarrow \infty$. 
This suggests that the two methods could be used together to cover virtually the full 
range $\eta/s$ parameter.
However for the phenomenological study of the QGP dynamics in ultra-relativistic collisions
both methods cover the relevant range usually explored, $0.08 \leq \eta/s \leq 0.2$.
The RBM-TP would be also the most suitable approach to study systems where
the physical conditions are such that one evolves from the low viscosity to the ballistic regime, as 
can occur in electron fluids in graphene constrictions \cite{Pellegrino:2017}.

We finally mention that the RBM-TP provides a solution of the Boltzmann equation for
the one-body distribution function $f( (x^{\alpha}), (p^{\alpha}) )$, allowing to evaluate  not only the 
$T^{\mu \nu}$ components, but several physical quantities, like
the transverse momentum spectrum, the azimuthal anisotropies $v_n$
as a function of the momentum, opening to a comparison with the wealth of experimental data.
However this requires to employ the output of the RBM-TP with a Cooper-Frye hypersurface and a 
statistical hadronization as a source for the multicomponent hadronic transport like SMASH \cite{weil-prc-2016}
or uRQMD, as done already within hydrodynamical approach \cite{petersen-jpg-2014,shen-cpc-2016}.

Furthermore in Ref. \cite{Plumari:2015sia} it has been show that is possible to evaluate the viscous
correction to the distribution function $\delta f(p)$, an essential quantity for the solution of the
Israel-Stewart viscous hydrodynamics.  Of course, the access to such a wider class of observables
comes at a price in computational time, that is currently more than two orders of magnitudes larger 
for RBM-TP w.r.t. RLBM.

%===================================================================================================
\section{Conclusions}\label{sec:conclusions}
%===================================================================================================

Summarizing, in this paper we have compared two different microscopic approaches, 
namely the RLBM, based on the simplified Anderson-Witting relaxation time approximation,
and RBM-TP, that by test particles method solves the full Boltzmann collision kernel. 
While these two models build on different methodologies,
it is quite remarkable that they yield to comparable results in the framework
of relativistic hydrodynamics. In particular, a comparison of the two methods
on a benchmark problem (the relativistic Riemann problem) shows matching
results. The simulations have been performed at different values of the
relativistic parameter $\zeta$, and under different viscous regimes (i.e with 
different values of the parameter $\eta/s$).

In the ultra-relativistic case, we have correctly compared our results against
the analytical solutions that are available in the inviscid and free-streaming
regimes, as well as against  data from a third numerical method,  BAMPS, for all values
of $\eta/s$ between these two limits. It is worth noting that the test particle
method is not suitable to simulate fluid flows in the limit $\eta/s \to 0$, since in this
region, numerical instabilities arise, due to unphysical values of the
particles cross sections. 
However the comparison with RLBM shows that it is already very reliable even at very low
viscosity such as $\eta/s \simeq 0.05$, smaller than the conjectured AdS/CFT lower bound.
On the other hand, RLBM begins to struggle at
values $\eta/s > 0.5$, where the hydrodynamic description starts to become
questionable, although still capable of reproducing the general behaviour of  the dynamics. 

This suggests that the two methods could be applied in
different zones of the parameter space, still featuring a window of
cross-compatibility {\it which permits their handshaking in prospective
multiscale simulations of quark-gluon plasmas}.
The specific case of the phenomenology of QGP in ultra-relativistic collisions
where $\eta/s(T)$ ranges  in $0.08<\eta/s < 0.2$, is indeed within the range of
cross-compatibility of the two methods.

After having established a firm connection to external results in the
ultra-relativistic limit, we have tested the two numerical schemes on
simulations at values of $\eta/s$, typical of quark gluon plasmas produced in
colliders like RHIC and LHC  and for several different values
of the rest mass of the particles. The two methods are in excellent agreement
when comparing the profiles of macroscopic quantities of interest, such as the
hydrostatic pressure and the macroscopic velocity. Some limited difference emerges when comparing the
non-equilibrium components of the particle four-current and the energy-momentum
tensor associated to heat-flux or dynamical pressure. 
For the future, would be interesting to explore whether the excellent agreement reported
in this paper still holds for initial profiles closer to the initial stage of ultra-relativistic collisions,
such as the Gubser \cite{gubser-prd-2010}  and the Bjorken profiles
\cite{bjorken-prd-1983}. 
The results presented in this paper allow to benchmark the results of newly developed numerical tools 
against two different numerical schemes; this process of cross-validation would permit to test the accuracy of such schemes, thus
paving the way to reliable and accurate simulations of real physical systems, such as quark-gluon plasmas dynamics in 
current and future high-energy experiments.
%===================================================================================================
\section*{Acknowledgment}
%===================================================================================================

DS has been supported by the European Union's Horizon 2020 research and
innovation programme under the Marie Sklodowska-Curie grant agreement No. 765048.
SS acknowledges funding from the European Research Council under the European
Union's Horizon 2020 framework programme (No. P/2014-2020)/ERC Grant Agreement No. 739964 (COPMAT). 
Numerical work has been performed on the COKA computing cluster at Universit\`a di Ferrara,
and  the computing cluster GR4-LNS at the INFN-LNS and the QGPDyn cluster at the{ University of Catania.

%===================================================================================================
% \section*{References}
%===================================================================================================

\bibliography{biblio}
%===================================================================================================

\end{document}

%===================================================================================================